\documentclass[aps,prl,twocolumn,color,psfig,epsf,floatfix]{revtex4-1}
\usepackage[english]{babel}
\usepackage[T1]{fontenc}
\usepackage{tikz}
\usepackage{amssymb,amsmath,amsfonts}
\usepackage{textcomp}
\usepackage{braket}
\usepackage{nicematrix} 
\usepackage{graphicx,color}
\usepackage[dvipsnames]{xcolor}
\usepackage[utf8x]{inputenc}
\usepackage{bm}
\usepackage{mathbbol}
\usepackage{float}
\usepackage{comment} 

\begin{document} 

\title{Geometry and topology of DNA defects} 
\author{A. Fassino$^1$, E. Carlon$^1$, D. Marenduzzo$^2$}
\address{$^1$ Soft Matter and Biophysics, KU Leuven, Celestijnenlaan 200D, 3001 Leuven, Belgium\\
$^2$SUPA, School of Physics and Astronomy, The University of Edinburgh, Edinburgh, EH9 3FD, United Kingdom}
    
\begin{abstract}
DNA defects arise frequently, either as a result of damage, 
protein binding or the action of molecular enzymes such as an RNA polymerase. 
Here we combine geometry, atomistic simulations and an analysis of real DNA structures to provide a quantitative characterisation of DNA defects and their dynamics. 
The composite order parameter describing defects geometrically is a backbone-to-base polar vector, together with a quaternion capturing 3D base orientation in a single strand. There are three types of defects: flips of the polar vector, stacking defects of bases, and composite defects encompassing both. 
All types of defects feature in realistic structures for DNA and have functional relevance. 
Additionally, biased molecular dynamics simulations of DNA with an imposed elastic deformation form flip or stacking defects whose abundance and dynamics strongly depend on sequence and type of deformation. 
Our results provide a link between geometry and biochemistry, and could also be applied, for instance, to identify defects in RNA molecules and nucleosome arrays.
\end{abstract}

\maketitle
 
Within living cells, the most frequently observed form of DNA is the B-form: its geometry is that of a right-handed double helix with approximately $10.5$ base pairs per pitch and a characteristic structural asymmetry that gives rise to the major and minor grooves~\cite{call97}. 
Departures from this idealised geometry are quantified in terms of relative translations -- slide, shift and rise -- and relative rotations -- roll, twist and tilt -- between neighbouring base pairs~\cite{call97,olso01}. For sufficiently small deformations, the associated elastic cost can be described within a harmonic approximation~\cite{bate05}, in close analogy with the Frank elastic free energy of liquid crystals, which quantifies the energetic cost of splay, bend and twist distortions~\cite{dege93}. This framework has proved highly successful for describing thermal fluctuations, sequence-dependent curvature and moderate supercoiling~\cite{mark94b}.

However, whilst textbooks depict DNA as a very stable molecule, in reality it is fragile, and can relatively easily undergo deformations that are too large to be regarded as small elastic perturbations of the B-form double helix~\cite{lind93}. Such configurations are more naturally interpreted as \emph{defects} of the helical structure. Common examples arise from DNA damage~\cite{frie03,cook01,frie05,sarm24,an14}, for instance due to oxydative lesions or UV ligth~\cite{poet20}. Damage often results in 
DNA lesions that induce strong local distortions of the double helix and may lead to base flipping, local unstacking and disruption of Watson-Crick pairing. 
Indeed, eukaryotic cells, such as human cells, protect DNA by wrapping it around proteins to form the chromatin fibre, which, besides inducing compaction,
has arguably evolved to protect the DNA mechanically from defect formation~\cite{call97}. 
   
Large deformations of DNA can also arise independently of chemical damage. For instance, protein binding 
frequently induces severe bending or local unwinding of DNA~\cite{call97}. 
Recent studies have characterized the free energy landscape of these highly bent (kinked) DNA conformations \cite{fass26}.
Similarly, highly transcribed genomic regions accumulate torsional stress and supercoiling~\cite{naug13,gilb25,naug26}, which can drive substantial local departures from canonical B-DNA and may nucleate defect-like conformations~\cite{mitc11,sutt16}. Despite their biological importance and frequent occurrence, such strongly deformed DNA states remain comparatively less studied from a geometric and theoretical perspective than small elastic distortions.  This is partly because the latter admit a natural description in terms of quadratic free-energy functionals, whereas large structural disruptions require a framework capable of describing discrete geometric transitions, including base flipping, local strand defects and strong 
reorientation of the base-pair frames.

In this work, we develop a geometric framework for describing large localised defects in the DNA structure, motivated by analogies with topological defects in soft condensed matter~\cite{merm79} and liquid crystals~\cite{dege93,lavr01,alex12,john25}. To do so, 
we represent each nucleotide by a local orientational frame and associated quaternion, allowing 
large discrete rotations and local flips to be characterised, both geometrically 
and topologically.
While we focus here on DNA defects, the introduced formalism is quite general and can be
applied to other problems, as for instance RNA or nucleosomal conformation in chromatin.

\begin{figure}
    \centering
    \includegraphics[width=\linewidth]{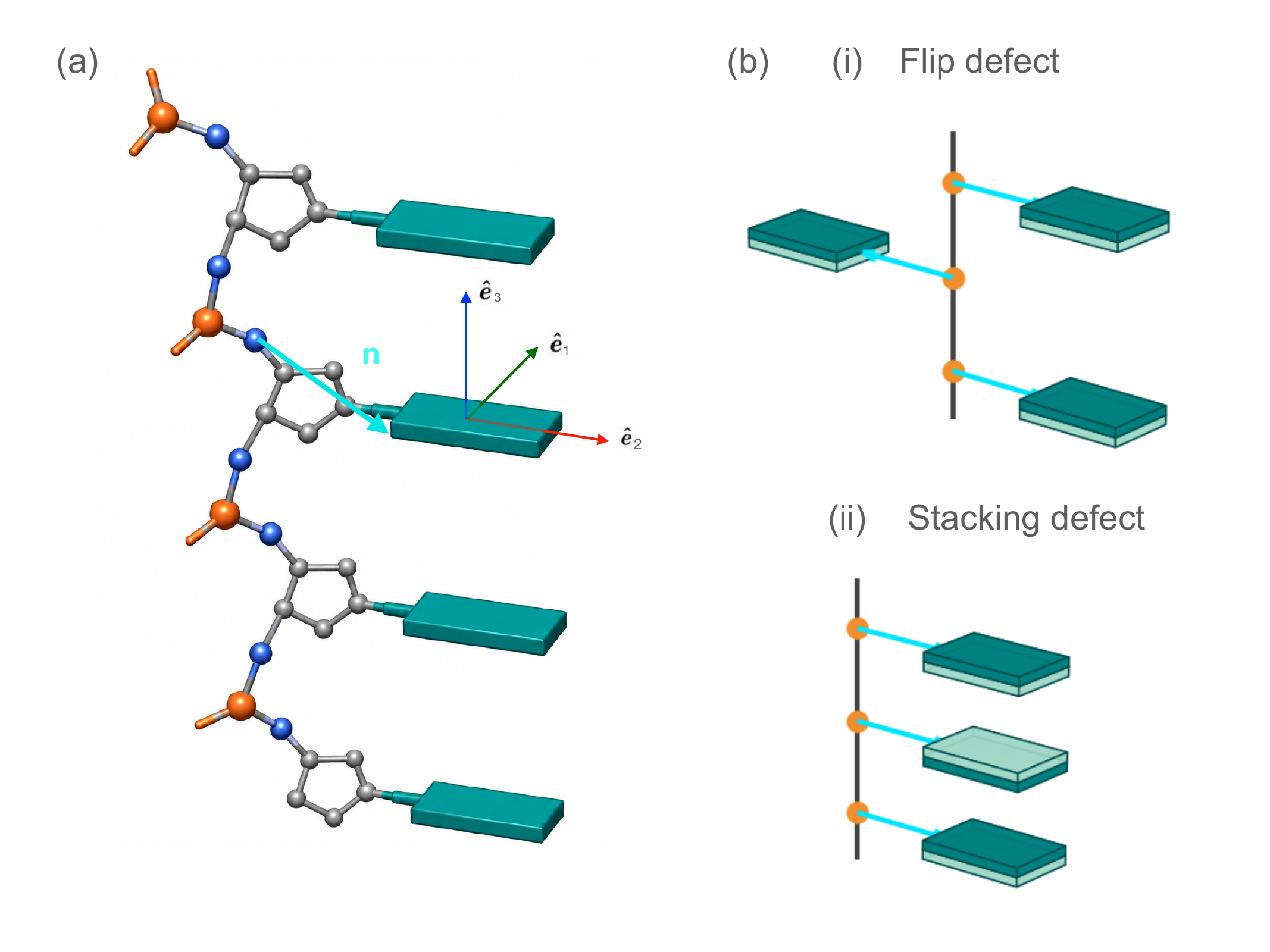}
    \caption{(a) Definition of ${\mathbf n(s)}$ and $q(s)$. (b)(i) Example of a flip defect, where the backbone-to-base vector flips. (ii) Example of a pure stacking defect, where the quaternion brick rolls by $2\pi$ in two steps of $\pi$. Combined flip and stacking defects are also possible in which (i) and (ii) both occur.}
    \label{fig1}
\end{figure}

\paragraph{Geometry and topology of DNA defects --} 
We model a single DNA strand as a 1D string, parametrised by $s$, with local orientation of a base described by two components [Fig.~\ref{fig1}(a)]. 
The first is the ``backbone-to-base vector'' $\mathbf{n}(s)$ which joins the $O5'$ oxygen to a nitrogen of the base pair, the $N1$ atom for a pyrimidine and the $N9$ for a purine~\cite{voor23}. The $\mathbf{n}(s)$ vector is approximately normal to the strand backbone unit tangent vector, ${\mathbf{t}(s)}$, hence the backbone-to-base 
vector can be parameterised by a single angle, and it corresponds to an element of $S^1$.
The second component is the base orientation, described as a rectangular ``brick'', 
which can be thought of as a biaxial nematogen, or a triad, and hence is equivalent to, 
and can be represented by, a unit quaternion 
$q(s)$, or a 3D rotation in $\mathrm{SO(3)}$. 
~\footnote{
It is useful to recall here that, because quaternions are equivalent to $\mathrm{SU(2)}$, they actually provide a double cover of the physical rotation space $\mathrm{SO(3)}$. This means that $q$ and $-q$ correspond to the same rotation in $\mathrm{SO(3)}$.} 
We will refer to the pair  $\left({\mathbf n}(s),q(s))\right)$ as the local order parameter describing base orientation along one DNA strand.

The local order parameter manifold describing a base on one of the strand is therefore 
${\mathcal M} = S^1 \times \mathrm{SO(3)}$. 
If we consider the case in which the initial and final states on one strand (on either 
side of a defect) are identical, for instance after a full turn of the double helix, 
or within a DNA loop, then these states are represented by the same 
element of 
${\mathcal M}$. Hence we can imagine joining the 
two ends so that the configurations can be topologically classified by the elements 
of the first homotopy group of ${\mathcal M}$, or $\pi_1({\mathcal M})$~\cite{merm79}, which we show in~\cite{SI} to equal $\mathbb{Z}\times \mathbb{Z_2}$.
Physically, the elements of $\pi_1({\mathcal M})$ represent classes of macroscopically distinct configurations that cannot be continuously deformed into one another without physically cutting the strand or completely flipping an 
orientation. These elements can be associated with a pair of integers $(m,n)$. Of these, $m$ denotes the winding of the vector along the strand under consideration~\cite{SI} 
while $n$ characterises  the 
topology of the triad describing the base, or the brick. The latter index 
corresponds to a spin-like twist parity of the triad, a topological property intuitively demonstrated by Dirac's belt trick~\cite{bolk73}.

While these global topological invariants classify all loop in order parameter space, here we are mostly interested in localised deformations or point-like defects. 
For these, $m$ can only in practice be $0$ or $1$~\cite{SI}. As a result we can classify each base step into one of four topological states. First, $(m,n)=(0,0)$ corresponds to a defect-free step. Second, $m=1$, corresponds to a``flip defect'', which is associated with a local reversal of the backbone-to-base vector [Fig.~\ref{fig1}(b)(i)]. Third, $n=1$ signals a ``stacking defect'', which corresponds to an abrupt $\pi$ rotation of the base triad [an example is shown in Fig.~\ref{fig1}(b)(ii)]. The latter rotation can be around the axis of the strand tangent, or the axis ${\hat{\mathbf{e}}_3}$ (quaternion swivel), around ${\hat{\mathbf{e}}_1}$ (quaternion roll), or around ${\hat{\mathbf{e}}_2}$ (quaternion flap). The three corresponding types of stacking defects are topologically equivalent and all have $n=1$; however, the attachment to the backbone means that only a $\pi$ quaternion roll is sterically possible without clashes. Swivel and flap stacking defects instead in practice also require a concurrent flip defect, so that for them $(m,n)=(1,1)$, a combined flip and stacking defect [for a roll stacking defect instead $(m,n)=(0,1)$ is possible]. 

We now discuss how ideal defects can be described by 
a $(\mathbf{n}(s),q(s))$ curve in ${\mathcal M}$, considering for simplicity 
a straight DNA backbone aligned with the $z$-axis (as in Fig. 1; so that $s=z$).  
For a flip defect [Fig.~\ref{fig1}(b)(i)], the backbone-to-base vector $\mathbf{n}(s)$ 
points radially outward (e.g., along $+\hat{\mathbf{x}}$) except at the defect location 
$s=s_0$, where the base is flipped by $\pi$:
\begin{equation}
\label{flipdefect}
\mathbf{n}(s) = \hat{\mathbf{x}}, \, {\rm if} \, s \neq s_0; \qquad \mathbf{n}(s_0)=-\hat{\mathbf{x}}.
\end{equation}
A (roll) stacking defect [Fig.~\ref{fig1}(b)(ii)] can instead be described by having 
the backbone along $z$, a constant ${\mathbf{n}}=\hat{\mathbf{x}}$, and a quaternion field given by:
\begin{equation}
\label{stackingdefect}
    q(s) = 1 \, \, {\rm if} \, s<s_0; \, \, q(s_0)={\mathbf{j}}; \, \,  q(s) = 1 \, \, {\rm if} \, s>s_0.
\end{equation}
An example of this pure stacking  example is Hoogsteen base pairing~\cite{hoog63}, where 
a purine rotates about its glycosidic bond to establish an alternative 
hydrogen-bonding pattern with the opposing pyrimidine. Watson–Crick/Hoogsteen 
interconversion occurs on $\mu s$–$ms$ timescales~\cite{niko11}. 
Formulas analogous to Eq.~(\ref{stackingdefect}) can be used to characterise quaternion swivel and quaternion flap, with ${\mathbf{j}}$ replaced by ${\mathbf{i}}$ or ${\mathbf{k}}$, although as mentioned these would also require a flip defect due to the geometry of the bases, which would usually cause a steric clash. Besides the true defects in Eqs.~(\ref{flipdefect},\ref{stackingdefect}), which cannot be continuously deformed to yield a defect-free molecule, 
below we will also track strong deformations which are also biologically relevant whilst not being true topological defects, 
for instance configurations where one base quaternion turns by $\pi/2$ and then relaxes back. 


To ensure that our classification of discrete local transitions (flips and stacking defects) reflects real geometrical disruptions of the bases rather than curvature of the DNA backbone, we also include a geometric correction based on parallel-transport (Bishop) frames~\cite{SI}.
When a single strand undergoes significant bending or writhe, consecutive nucleotides exhibit changing orientations within the global (laboratory) reference frame. This macroscopic curvature introduces a 
geometric contribution to the raw base-step vector dot products and quaternion transitions, which can mask defects. By projecting both the polar backbone-to-base tracking vectors $\mathbf{n}(s)$ and the rigid-body triads into a parallel-transported Bishop frame constructed along the $O5'$ backbone centerline trajectory, we  then 
factor out the spatial variations induced by 3D writhing -- details of this procedure, which we have used in the following, are given in~\cite{SI}. 



\paragraph{Naturally occurring DNA defects --}
Conformational defects in naturally occurring DNA structures arise primarily through two mechanisms: intrinsic thermal fluctuations and perturbations induced by external factors such as torsional stress (e.g., transcriptionally-induced supercoiling~\cite{naug13,naug26}) or protein binding. In some cases, protein defects are also signals which promote binding of repair proteins~\cite{ghed25,sara25}.
Defects arising instead from thermal fluctuations are typically transient local excitations of the DNA duplex, corresponding to 
short-lived departures from the canonical B-form geometry, which relax back to the native state. 
Although spectroscopic techniques can provide information on the populations and lifetimes of these transient excitations, their detailed atomic dynamics remain inaccessible to direct observation. In contrast, X-ray crystallography and cryo-EM structures deposited in the Protein Data Bank (PDB)~\cite{pdb} provide high-resolution 
snapshots of DNA conformations, including locally distorted states. 

\begin{figure}
    \centering
    \includegraphics[width=\linewidth]{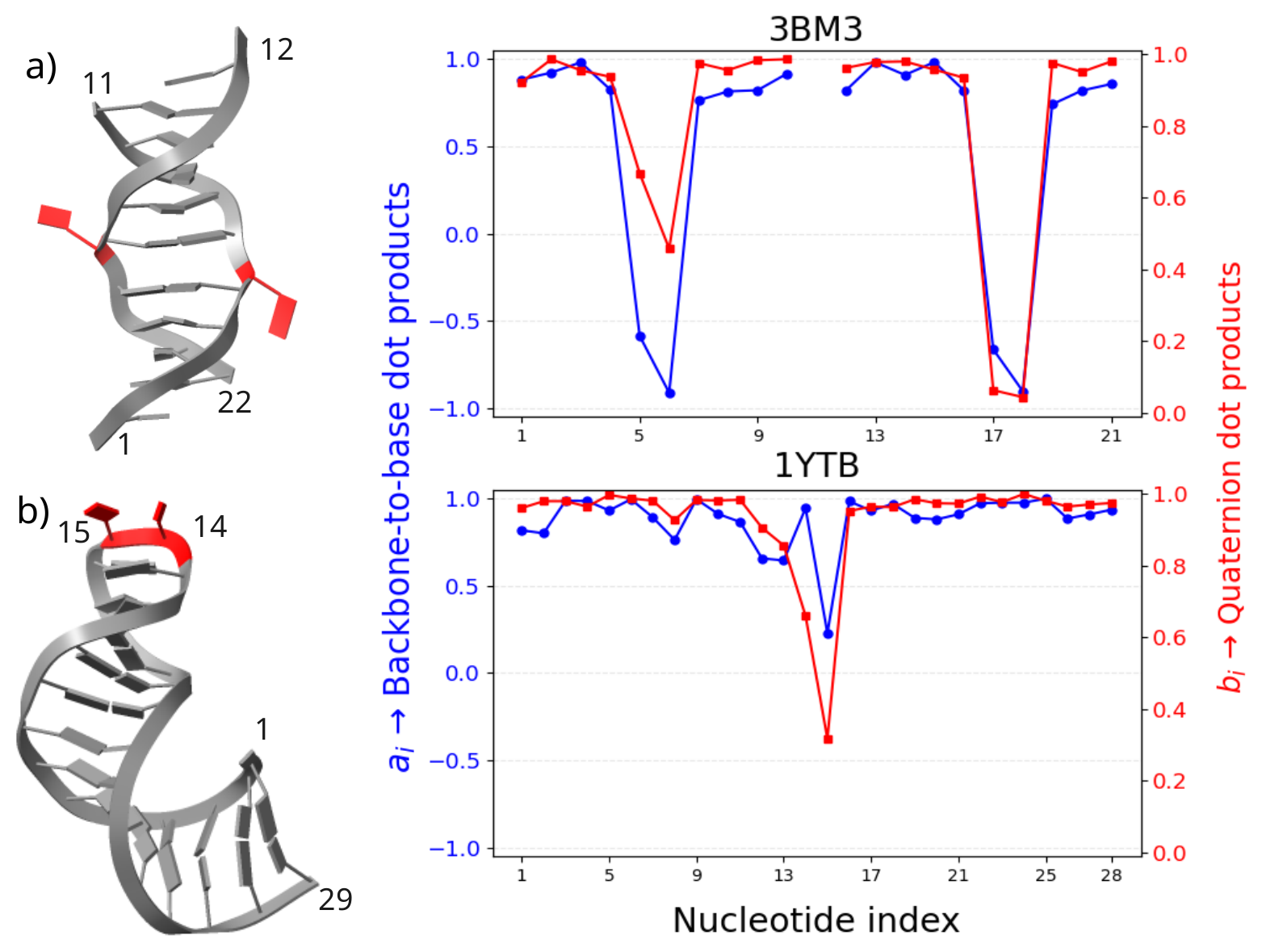}
    \caption{(a) Example of a DNA molecule with two flip defects (PDB entry:3BM3). One of the
    flip defect is also a stacking defect. (b) Example of 
    a stacking defect (1YTB). In the snapshots defects are highlighted in red.
    }
\label{fig2}
\end{figure}

An example of PDB structure of local defects induced by protein binding (Type II restriction endonuclease) is given in Fig.~\ref{fig2}, left. The protein-DNA binding gives rise to a base-flipping defect in each DNA strand~\cite{szcz08}. 
The right panel plots the values of two order parameters which are the dot products of two consecutive backbone-to-base vectors $a_i \equiv \hat{n}_i \cdot \hat{n}_{i+1}$ (blue) and of two consecutive base quaternions $b_i \equiv |q_i \cdot q_{i+1}|$ (red). 
The index $1 \leq i \leq 9$ labels bases on strand A, while $10 \leq i \leq 21$ those on strand B, as illustrated in Fig.~\ref{fig2}(a), left.
We note that a single base defect at  base $i$ results in deviations of the two neighboring order parameters $a_{i-1}$, $a_i$, as well as $b_{i-1}$, $b_i$. Our geometric analysis reveals that the defect on both strands corresponds to a nearly pure inversion of the backbone-to-base polar vector as $a_{i-1} \approx a_i \approx -1$. The degree of unstacking is however different in the two strands with the defect in strand B significantly more unstacked ($b_{i-1} \approx b_i \approx 0$) as that in strand A ($b_{i-1} \approx b_i \approx 0.5$). 
The structure of Fig.~\ref{fig2}(b) shows a more modest flipping and weak 
unstacking of a single base pair ($a_{i-1} \approx 1$, $a_i \approx 0$ and 
$b_{i-1} \approx 0.6$, $b_i \approx 0.3$).
These protein-induced distortions provide experimentally resolved examples of conformational defects and offer valuable structural benchmarks for characterising non-canonical DNA states. 

\begin{figure*}[!th]
    \centering
    \includegraphics[width=\textwidth]{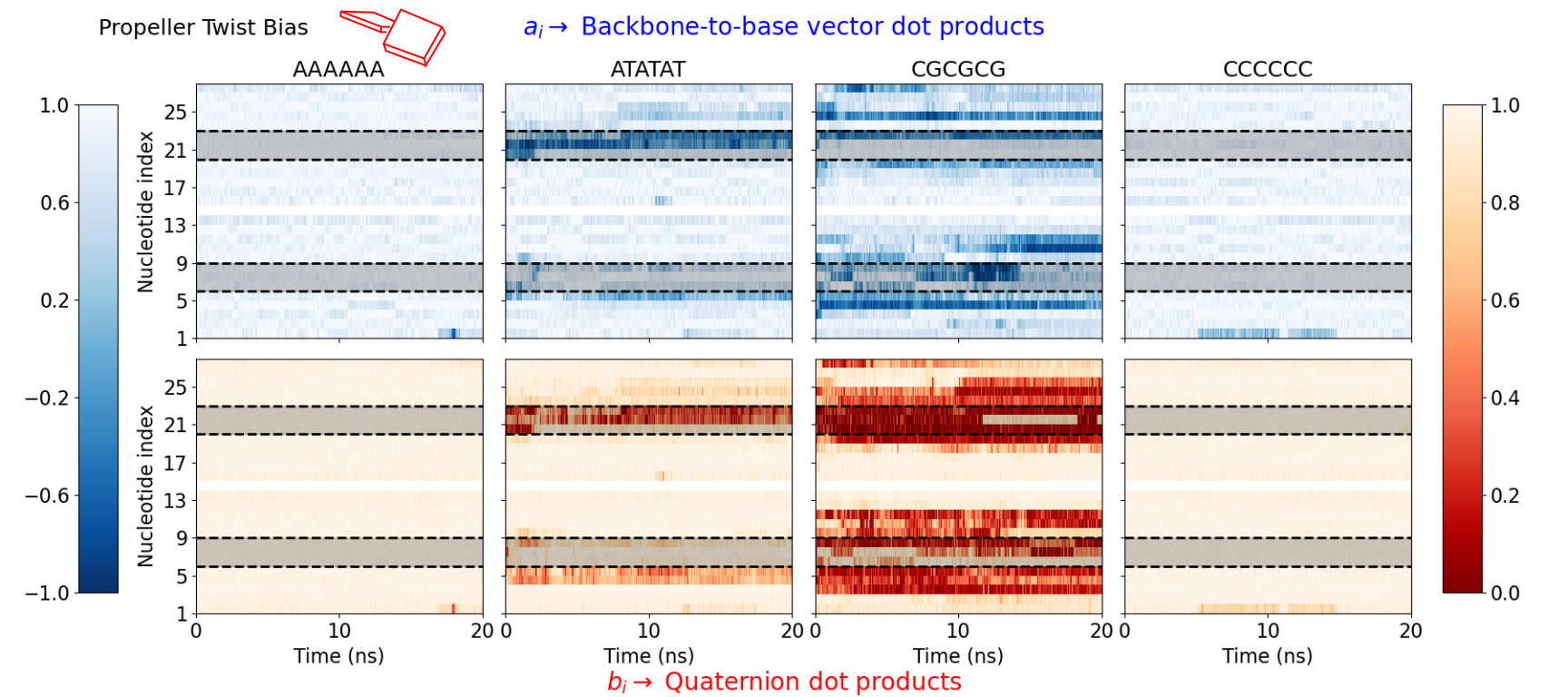}
    \caption{Kymograph showing deformations in the backbone-to-base and quaternion orientations, corresponding to simulations with a local bias applied to propeller twist in the middle of a DNA 14-mer. The bias was applied for $1$ ns, with $\bar{\alpha} =0.3$ (not shown in the Figure); then the system was let to equilibrate with the bias removed for $20$ ns (shown in the Figure). The region where the bias is applied is shown within dashed lines in the kymographs.
}
    \label{fig:kymograph}
\end{figure*}

\paragraph{DNA defects in atomistic simulations --}
Experimental structures from PDB provide static snapshots of DNA and hence offer limited insight into the dynamical pathways by which conformational defects nucleate and propagate. To investigate both the propensity for defect formation of a given DNA sequence and their subsequent dynamics, we perform all-atom molecular dynamics 
simulations (see~\cite{SI} for more details on the protocol used). 
Briefly, to induce defect formation, we impose local conformational biases using the rigid base biasing for nucleic acids (RBB-NA) algorithm~\cite{voor23}. RBB-NA adds a harmonic biasing potential acting on a coarse-grained rigid-base coordinate $\alpha$ of the type 
\begin{equation}
V_{\rm bias}(\alpha)
= \frac{K}{2}\left(\alpha-\bar{\alpha}\right)^2,
\label{def:bias}
\end{equation}
where $K$ is the stiffness of the restraint and $\bar{\alpha}$ is the target value of the coordinate which we consider. The latter 
can be any of the twelve rigid-base parameters~\cite{olso01}, such 
as twist, roll, propeller twist, opening or buckle. The bias \eqref{def:bias} drives $\alpha$ toward the prescribed $\bar{\alpha}$. The RBB-NA algorithm has previously been employed in umbrella-sampling studies to characterize the free-energy landscape of strongly bent and/or twisted DNA conformations~\cite{fass26}. Here, we use the same approach to induce controlled departures from canonical B-DNA by applying sufficiently strong restraints, i.e., large values of both $K$ and $\bar{\alpha}$. 
With this method, 
we can investigate some key aspects of defect formation and propagation in DNA. First, we can quantify sequence-dependent effects, identifying for instance sequence motifs that are more or less susceptible to defect formation. Second, 
our simulations provide access to the spatial extent and timescales over which these perturbations develop, relax, and propagate along the DNA molecule.  

Fig.~\ref{fig:kymograph} shows kymographs depicting the temporal evolution of $a_i$ and $b_i$ throughout the DNA molecule over the timescale of $20$~ns of unbiased simulation for four different 14-mer sequences: polyA, polyAT, polyCG and polyC subjected to an initial propeller twist bias in the middle base pair~\footnote{The sequence is CGCGAAAAAACGCG, so 6 polyA surrounded by CG. The same holds for the other sequences~\cite{SI}.}. The main feature of the results is that the sequence strongly affects the likeliness of forming structural defects both at the initially biased region and elsewhere. Specifically, PolyA and PolyC sequences are virtually defect free. 
Instead, PolyAT and PolyCG are strongly deformed. For PolyCG, contiguous stacking defects proliferate throughout the 14-mer, whereas flip defects, also frequent, are isolated and usually non-contiguous. For PolyAT, combined flip and stacking defects  form at a distance from the biased base pair, towards the middle of the molecule. 

These results are consistent with {\it in vivo} experiments showing that gene promoters in CpG islands are easier to melt especially when subjected to negative supercoiling from polymerase activity~\cite{naug26}, notwithstanding the fact that CG base pairs should be more thermodynamically stable to melting than AT base pairs~\cite{call97}. They are also in line with single molecule experiments showing that CG-rich sequences can melt more easily than AT-rich ones, again when subjected to torsional stress~\cite{vlij2015}. Additionally, we argue that the defect formation at a distance observed in PolyAT provides a functionally relevant example in which deformation signals at a base pair (propeller twist in Fig.~\ref{fig:kymograph}) can mediate long-range communication through DNA, giving rise to effective interactions between proteins bound several base pairs apart, and thereby contributing to allosteric mechanisms of DNA recognition and regulation~\cite{sege23}.

\begin{figure}
    \centering
    \includegraphics[width=\linewidth]{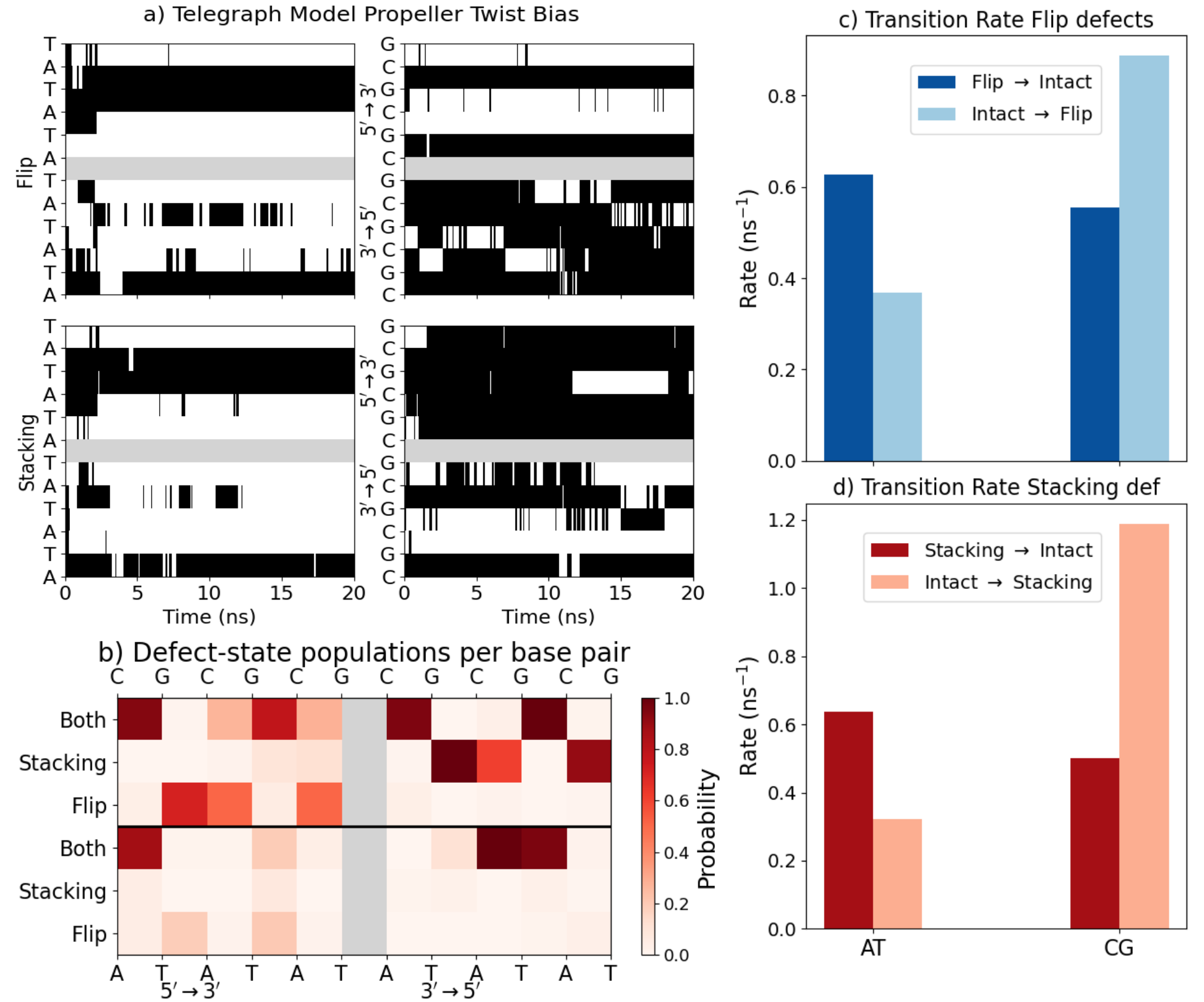}
    \caption{Dynamics and statistic of defects under a propeller twist bias derived from Hidden Markov Model analysis. (a) Representative telegraph model trajectories tracking defect states over $20$ ns for PolyAT and PolyCG sequences (black intervals denote defect presence). (b) Probabilities of finding specific defect states -- pure flip, pure stacking, or combined ("Both") -- per base pair. (c, d) Forward and backward transition rates between intact base pairs and (c) flip defects or (d) stacking defects, for AT and CG steps.}
    \label{fig:propeller_analysis}
\end{figure}

The kymographs in Fig.~\ref{fig:kymograph} suggest that the dynamics of flip and stacking defects is 
strongly sequence-dependent. To quantify the kinetics of defect formation and annealing, we mapped the time-dependent trajectories of $a_i$ and $b_i$ to a telegraph model~\cite{gard85} using a Gaussian Hidden Markov Model [HMM; see Fig.~\ref{fig:propeller_analysis}(a), and End Matter for details]. Each nucleotide is therefore mapped to one of four states: no defects, flip defect, stacking defect, and combined (flip plus stacking) defect. 

The analysis of the HMM tracks occupation statistics [Fig.~\ref{fig:propeller_analysis}(b)] shows that, under a propeller twist bias, PolyAT 
displays predominantly combined defects, whereas PolyCG defects show either 
combined defects or stacking defects, both of which are often clustered. 
The order of magnitude formation and annealing rates is $\sim$ ns$^{-1}$ for both flip and stacking types [Fig.~\ref{fig:propeller_analysis}(c,d)], and the rate of formation is significantly larger for PolyCG.
Finally, the probability distributions for defect annealing times [the ``defect survival probability'' in Fig.~\ref{fig:survivalprobability}(a), End Matter] is not exponential in the propeller twist bias simulations. This suggests that there is not a single timescale for defect healing, as 
correlated defect islands commonly appear, and the timescale for annealing a defect island depends on its size. 

Simulations with buckle and opening bias (see End Matter) confirm the qualitative trends seen for the propelled twists, that defect formation and annealing depends on sequence and is $\sim$ ns$^{-1}$. However, they also show that the geometry of the imposed bias strongly contributes: for instance, an opening bias leads to a panoply of defects in PolyAT sequences and very few for PolyCG ones, contrarily to the propeller twist bias. 
  
\paragraph{Conclusion}
In summary, we have proposed a geometric framework for the characterisation of large, 
localised structural perturbations in DNA, by using concepts from defects in soft matter physics and from topology~\cite{merm79}. By introducing 
a composite order parameter consisting of a polar backbone-to-base vector $\mathbf{n}(s)$ and a unit quaternion $q(s)$ 
corresponding to the brick-like geometry of a DNA base, we have shown that DNA defects 
can be classified into three types: flip defects, stacking defects, and combined defects which comprise both. 

First, by analysing data in the Protein Data Bank, we showed that DNA defects 
are often found {\it in vivo}, where they are functionally relevant. For instance, restriction enzymes exploit flip defects to isolate specific target sequences within their active cleavage sites, while stacking defects may act 
mechanical signals to be read by transcription factors.  
We have also shown evidence from biased atomistic molecular dynamics simulations that defect formation and annealing under elastic constraints is strongly sequence-dependent. Notably, alternated sequences (polyAT and polyCG) show a significantly heightened susceptibility to defect formation compared to homopolymeric tracts (polyA and polyC), with CpG steps being the most fragile under propeller twist stress, which may arise in highly expressed genes. 
This 
mechanical fragility of CpG regions 
could provide biological insight into the enriched formation of melted and non-B-DNA structures in these regions~\cite{naug26}. 
Our simulations also show that defect formation and annealing is a spatiotemporally correlated process strongly affected by the formation of  defect islands.

Although we focused on structural defects in bare DNA, these order parameters may also apply to other biological systems, such as chromatin, where DNA wraps around histone octamers to form nucleosomes. Because nucleosomes have well-defined geometries and orientations, quaternion fields could describe the relative orientations of consecutive nucleosomes, while flip vectors could quantify their orientation relative to the linker DNA. Therefore, composite order parameters such as that used here could provide a quantitative framework to identify defects at the nucleosome-array level~\cite{garc15}.

\paragraph{Acknowledgements --}
A.F. acknowledges financial support from the EU-funded Doctoral Network MeChaNiSM (Mechanical Characterization of Nucleic acids using Single Molecule techniques) within the Horizon Europe 2021–2027 Framework Programme Grant Agreement number 101168851. The resources and services used in this work were provided by the VSC (Flemish Supercomputer Center), funded by the Research Foundation Flanders (FWO) and the Flemish Government.

\bibliography{references}

\clearpage

\appendix
\setcounter{equation}{0}
\renewcommand{\theequation}{A\arabic{equation}}

\section{END MATTER}

\subsection{DNA defects in atomistic simulations with buckle and opening bias}

\begin{figure}[H]
    \begin{minipage}{\textwidth}
        \centering
        \includegraphics[width= \textwidth]{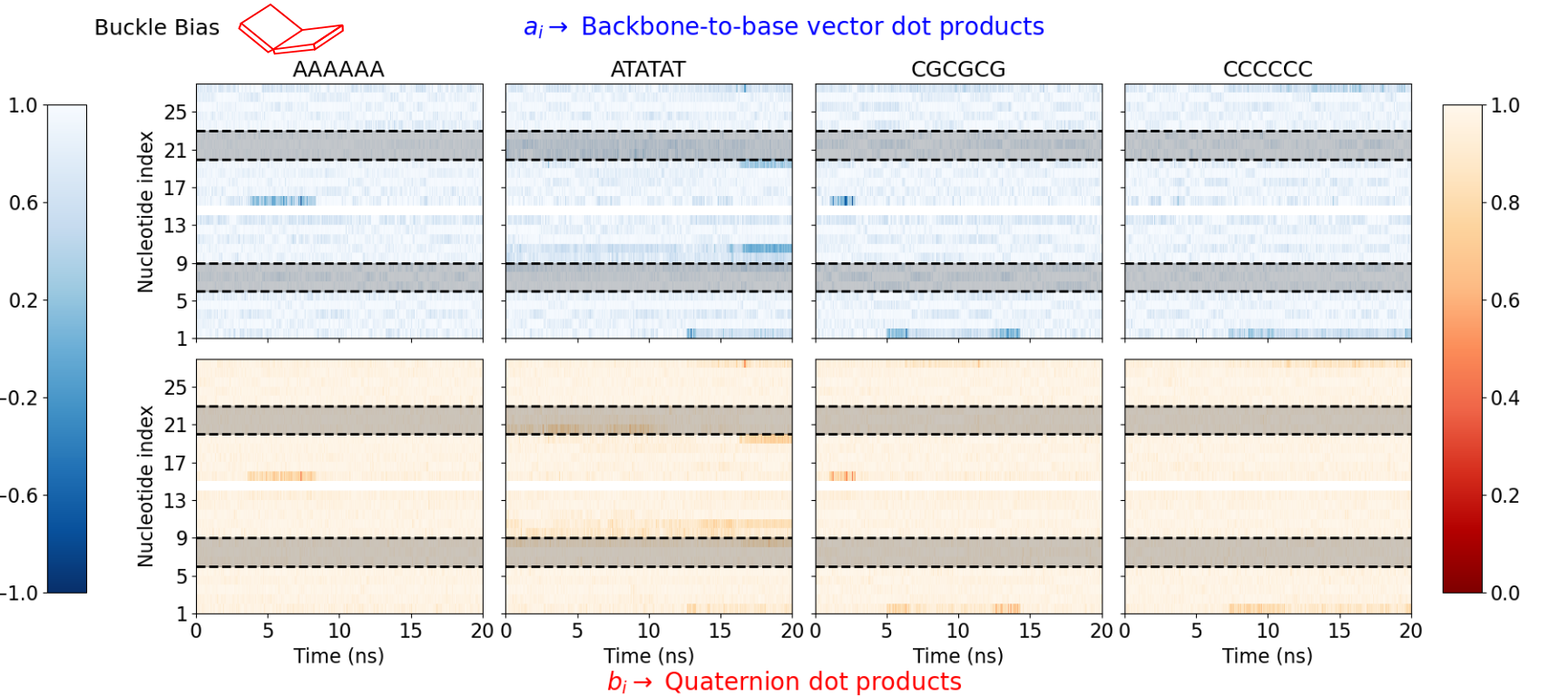}
        \caption{Kymograph corresponding to a $20$ ns simulation following $1$ ns with buckle bias equal to $\bar{\alpha} = 0.3\text{ rad}$.}
        \label{fig:bucklekymograph}
    \end{minipage}
\end{figure}

In the main text we showed results for atomistic simulations with propeller twist bias. 
In this Section, we discuss defect dynamics with other types of elastic perturbation. First, Fig.~\ref{fig:bucklekymograph} shows the kymograph corresponding to the same protocol used in Fig.~\ref{fig:propeller_analysis}, but for a buckle bias. The results show that a buckle perturbation leads to significantly fewer defects with respect to the propelled twist bias. The PolyAT sequence is the one which leads to more defects in this case. As for the propeller twist bias, there is action at a distance, as defects form a few base pairs away from the initially imposed perturbation. 

\onecolumngrid
\twocolumngrid

\onecolumngrid
\begin{figure}[H]
    \begin{minipage}{\textwidth}
    \centering
    \includegraphics[width=\textwidth]{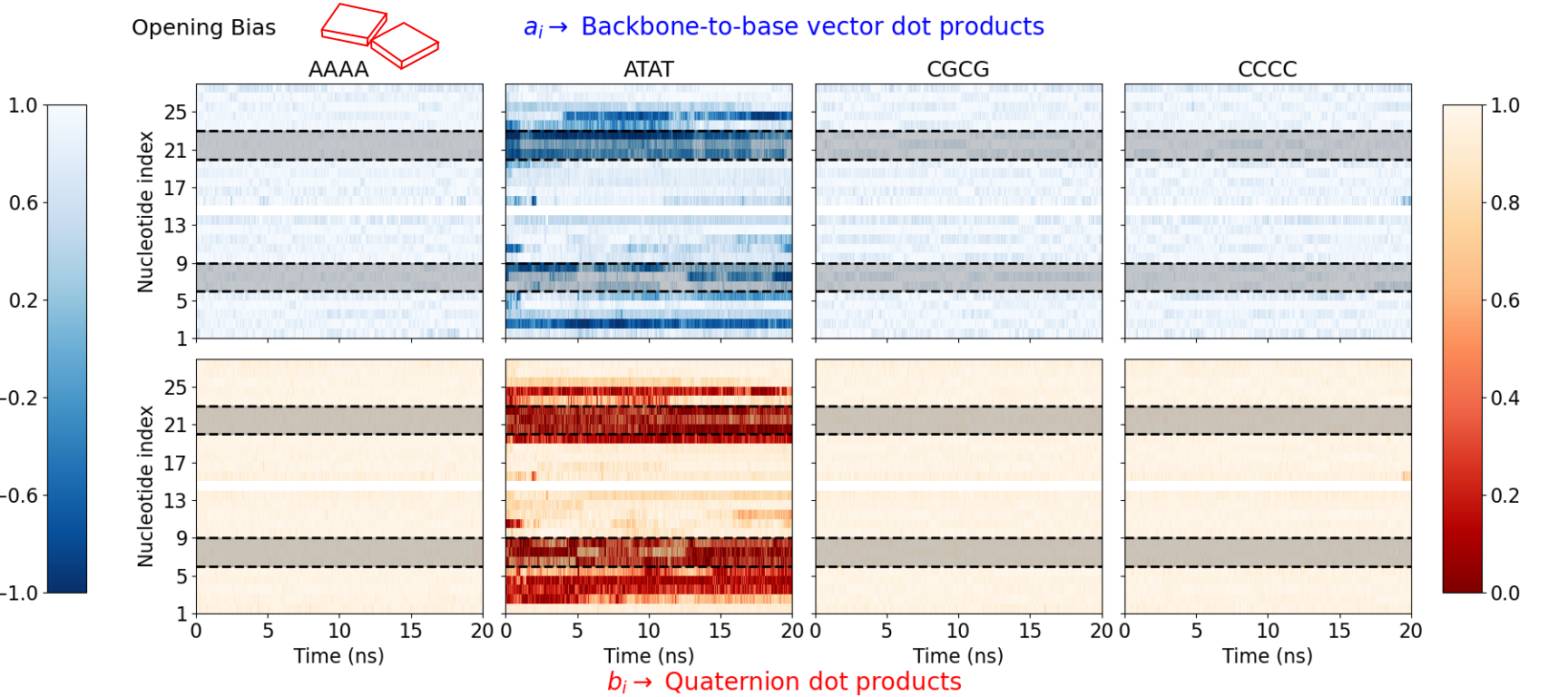}
    \caption{Kymograph corresponding to a $20$ ns simulation following $1$ ns simulation with opening bias equal to $\bar{\alpha} = 0.42\text{ rad}$.}
    \label{fig:openingkymograph}
    \end{minipage}
\end{figure}
\twocolumngrid

Fig.~\ref{fig:openingkymograph} instead shows the kymograph for an opening bias in the middle of the molecule. This bias creates more defect, similar to the case of propeller twist. Here, again the PolyAT chain is the more distorted, and stacking defects are the most common type of deformations. 

The propeller twist bias results in the main text showed that PolyAT is more compliant than PolyCG, as the former forms isolated defects more easily, whilst the latter favours stress accumulation which results in larger scale structural faults. With opening bias, instead, it is the PolyAT which forms more defect: we interpret this as due to the fact that AT are kept together by two rather than three H bonds, and opening couples more to the disruption of these H bonds. On the other hand, propeller twist disrupts more stacking, whilst buckle affects both H bonds and stacking less, and can therefore be   absorbed more easily in a DNA double helix. Therefore our results show that defect formation depends on both sequence and applied perturbation in a nontrivial way.

\begin{figure}[t!]
    \centering
    \includegraphics[width=\linewidth]{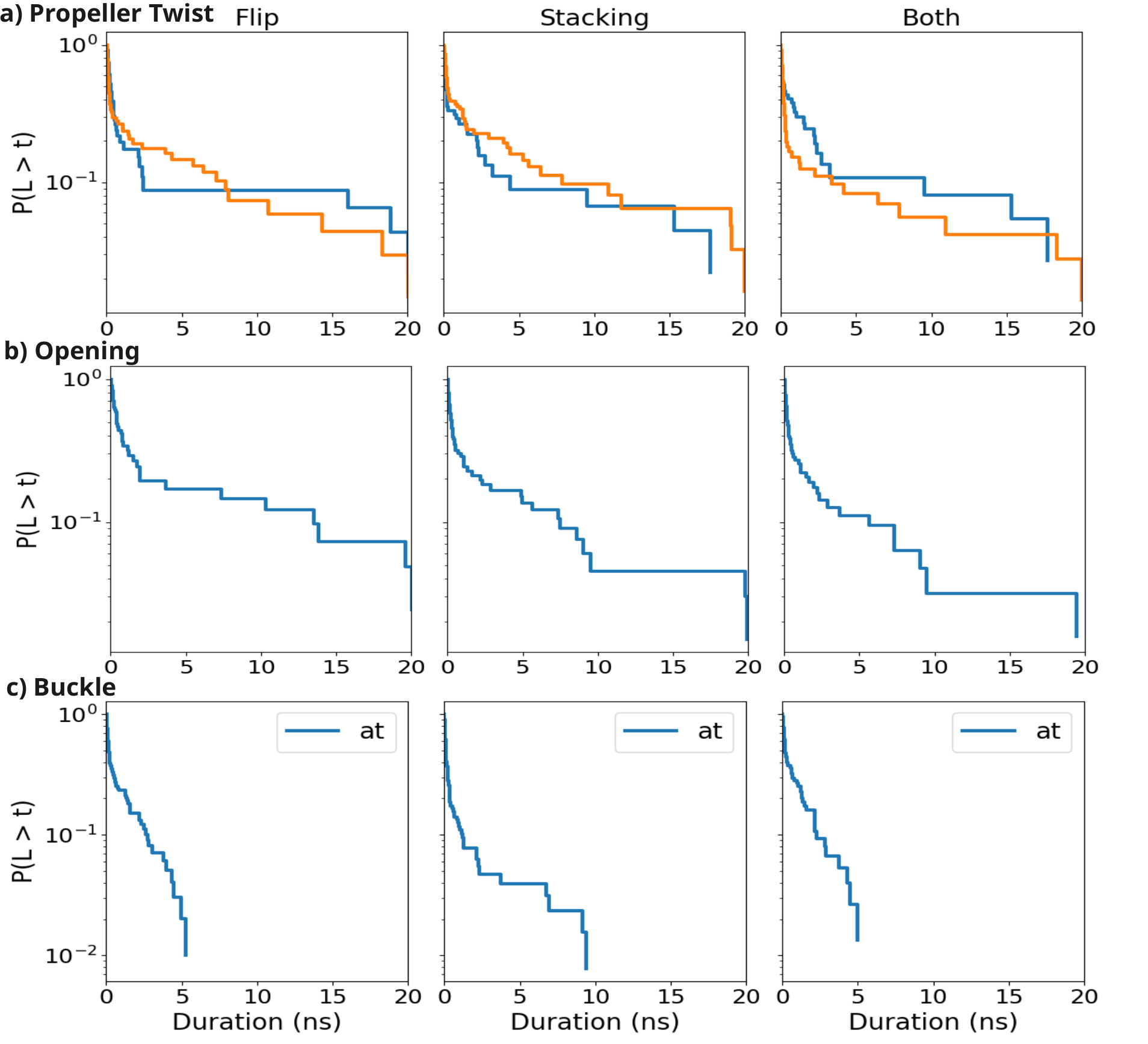}
    \caption{Survival probability for flip defects (left), stacking defects (centre) and combined flip+stacking defects (right). The three rows show the results from simulation with a bias on the propeller twist (top row), opening (middle row) and buckle (bottom row).}
    \label{fig:survivalprobability}
\end{figure}

\subsection{Hidden Markov model for DNA defects}

In Fig.~\ref{fig:propeller_analysis}, the populations of the two states were estimated using a Hidden Markov Model (HMM) algorithm. The model found the mean value of flip defects for Propeller Twist bias at $a_i=\cos\theta = 0$ and the mean value associated with thermal fluctuations at $\cos\theta = 0.8$. Depending on the stacking sequence, the characteristic values of $b_i$ for a step are approximately $b_i=0.36$ for AT stacking, $0.20$ for CG stacking, and $0.6-0.8$ for the canonical (normal) conformation.
We identified the two states using a Gaussian Hidden Markov Model (HMM) implemented in Python with the GaussianHMM class from the hmmlearn library. First, we tested hidden Markov models with $2$, $3$ and $4$ hidden states. In most cases, the likelihood scores indicated that a two-state model was sufficient to capture the underlying dynamics. In particular, for all stacking defects and for all flip defects in PolyAT, the likelihood obtained with the three-state model was essentially identical to that of the two-state model, indicating that the additional hidden state was unnecessary. The only exception is the flip defect in PolyCG under a Propeller Twist bias. In this case, the likelihood score for the three state model is substantially higher, suggesting the presence of three distinct hidden states, with mean values at approximately $\cos\theta=-0.2$, $\cos\theta=-0.5$, and $\cos\theta = 0.9$, the latter corresponding to thermal fluctuations around the equilibrium configuration. However, to maintain consistency across all analyses, we adopted a two state model throughout, corresponding simply to the presence or absence of the defect. The models were trained separately on the flip and stacking values of the $6$ central base pairs, corresponding to PolyAT and PolyCG, over the full $20$ ns trajectories considering frames every $10$ ps. From the result of the model we then assigned a state to each observation producing the telegraph plot and later determine the transition probability matrix. The state with lower mean value is identified as defect state, while the higher mean value is the non defective state. From the self transition probabilities $P_{ii}$ we calculated the residence times of each state $\tau_i=-\Delta t/\ln(P_{ii})$, and then determined the transition rates as $1/\tau_i$.

\subsection*{Survival probabilities and defect lifetimes}

Once flip and stacking defects are identified through the HMM, we can quantify the timescales for defect formation and annealing. Besides the rates discussed in the main text, it is useful to analyse the defect survival probability, or the distribution function for the lifetime of flip, stacking and combined defects. If defect formation and annealing were Poisson processes with a constant rate, these survival probabilities should be single exponential. Instead, Fig.~\ref{fig:survivalprobability} show that for all biases and defect types the curves are never single exponential. This shows that the relaxation of defect dynamics occurs at multiple time scales, due to correlations which create defect islands of different sizes. Because the annealing of a defect island depends on its size, multiple timescales appear naturally in the problem. When defects are persistent, as in PolyCG sequences under propeller twist bias, the lifetime of defects can extend to the duration of our simulations, which is consistent with the expectation that they persist for a sufficient time to be scanned by DNA-binding protein or the repair machinery~\cite{sara25}. 


\clearpage

\setcounter{figure}{0}

\renewcommand{\thefigure}{S\arabic{figure}}

\setcounter{equation}{0}

\renewcommand{\theequation}{S\arabic{equation}}

\setcounter{page}{1}    

\setcounter{figure}{0} 

\renewcommand\thefigure{S\arabic{figure}}  

\setcounter{equation}{0}

\renewcommand{\theequation}{S\arabic{equation}}

\setcounter{table}{0}

\onecolumngrid

\appendix

\section*{Supplemental Material: Geometry and topology of DNA defects}

\renewcommand{\thetable}{S\Roman{table}}

In this Supplemental Material we discuss how to compute the first homotopy group $\pi_1({\mathcal M})$, how to correct for strand curvature in our defect analysis pipeline, and what protocol we used to generate our atomistic simulation trajectories.

\section*{Homotopy group for the order parameter manifold}

 The composite order parameter describing a DNA base pair, $({\mathbf n}(s),q(s))$ lies on the manifold $\mathcal{M} = S^1 \times SO(3)$, which is a Cartesian product. The fundamental group of a product space factorises into the direct product of the individual fundamental groups, $\pi_1(S^1 \times SO(3)) \cong \pi_1(S^1) \times \pi_1(SO(3))$~ \cite{merm79}. Therefore, the task of finding $\pi_1({\mathcal M})$ reduces to that of computing $\pi_1(S^1)$ and $\pi_1(SO(3))$.
 
First, we can parametrise the backbone-to-base vector, which is perpendicular to the strand containing the base, via an angle in $S^1$, which here we call $\phi(s)$. Closed loops $\gamma$ are classified by an integer winding number \cite{merm79},
\begin{equation}
m = \frac{1}{2\pi} \oint_{\gamma} d\phi, 
\end{equation}
so that $\pi_1(S^1) \cong \mathbb{Z}$.

Similarly, for the rotation group $SO(3)$ describing the base triad, unit quaternions $q \in S^3$ double-cover $SO(3)$ (with $q$ and $-q$ representing the same rotation)~\cite{merm79}. Because $S^3$ is simply connected, a single $2\pi$ rotation around a fixed axis traces a path from $+1$ to $-1$ in $S^3$ and therefore defines a non-contractible loop in $SO(3)$, whereas a $4\pi$ rotation closes back to $+1$ and can be shrunk to a point. Hence $\pi_1(SO(3)) \cong \pi_1(S^3/\mathbb{Z}_2) \cong \mathbb{Z}_2$.

Combining both factors yields 
\begin{equation}
\pi_1(\mathcal{M}) \cong \mathbb{Z} \times \mathbb{Z}_2, 
\end{equation}
with homotopy classes labelled by $(m, n) \in \mathbb{Z} \times \mathbb{Z}_2$. Whilst in principle $m$ is an integer winding number, in practice a flip defect corresponds to two local $\pi$ reversals of the backbone-to-base vector, so that for application to DNA defects $m$ can either be $0$ (no flip defect) or $1$ (a flip defect).

\section*{Bishop frames and curvature correction for defect identification}

\subsection*{Relation between the Frenet and the Bishop frame}
The Bishop frame (also known as the parallel-transport frame) provides a continuously moving reference system along a three-dimensional space curve $\mathbf{r}(s)$ representing the DNA backbone trajectory. Unlike the standard Frenet-Serret formulation, which exhibits geometric singularities wherever the local curvature vanishes ($\kappa = 0$) or undergoes inflection, the Bishop frame minimises internal twisting by imposing a parallel-transport condition on the normal plane. 

Let the trajectory of the DNA backbone be parameterised by its arc length $s$, yielding a unit tangent vector field:
\begin{equation}
\mathbf{T}(s) = \frac{d\mathbf{r}}{ds}
\end{equation}
The Bishop frame defines two normal vectors, $\mathbf{N}_1(s)$ and $\mathbf{N}_2(s)$, which span the plane perpendicular to $\mathbf{T}(s)$, yielding a right-handed orthonormal basis $\{\mathbf{N}_1, \mathbf{N}_2, \mathbf{T}\}$. The characteristic feature of this system is that its normal vector fields do not rotate around the tangent vector field (there is zero relative torsion), or mathematically the variation of the basis vectors with curvilinear abscissa along the DNA strands obey the following differential equations:
\begin{equation}
\frac{d\mathbf{N}_1}{ds} = -\kappa_1 \mathbf{T}, \quad \frac{d\mathbf{N}_2}{ds} = -\kappa_2 \mathbf{T}, \quad \frac{d\mathbf{T}}{ds} = \kappa_1 \mathbf{N}_1 + \kappa_2 \mathbf{N}_2
\end{equation}
where $\kappa_1$ and $\kappa_2$ are the components of the normal curvature vector field. 

To track the internal molecular orientation spatially, the base triad matrix $R = [\mathbf{e}_1 \mid \mathbf{e}_2 \mid \mathbf{e}_3] \in SO(3)$ is mapped to a unit quaternion $\mathbf{q} = (x, y, z, w)$ through the Rodrigues rotation formula. The columns of the rotation matrix correspond to the orthogonal unit vectors $\mathbf{e}_1, \mathbf{e}_2, \mathbf{e}_3$:
\begin{equation}
\mathbf{e}_1 = \begin{bmatrix} 1 - 2(y^2 + z^2) \\ 2(xy + wz) \\ 2(xz - wy) \end{bmatrix}, \quad
\mathbf{e}_2 = \begin{bmatrix} 2(xy - wz) \\ 1 - 2(x^2 + z^2) \\ 2(yz + wx) \end{bmatrix}, \quad
\mathbf{e}_3 = \begin{bmatrix} 2(xz + wy) \\ 2(yz - wx) \\ 1 - 2(x^2 + y^2) \end{bmatrix}
\end{equation}
This parameterisation allows the tracking of local orientations in a numerically efficient way. 

\subsection*{Relation between quaternion and triad}
A unit quaternion and a triad have the same number of degrees of freedom ($3$), and they are physically equivalent. It is convenient to give some specific formulas which allow one to convert between triad and quaternion representations, as quaternions are more tractable numerically but triads are physically intuitive. 

To express the unit quaternion $\mathbf{q}$ in terms of the base triad vectors $\{\mathbf{e}_1, \mathbf{e}_2, \mathbf{e}_3\}$, we note that it corresponds to a spatial rotation about a unit axis $\hat{\mathbf{u}}$ by a physical rotation angle $\theta$, so that
\begin{equation}
\label{quaternioneq}
\mathbf{q} = \cos\left(\frac{\theta}{2}\right) + \sin\left(\frac{\theta}{2}\right) \left( u_x \mathbf{i} + u_y \mathbf{j} + u_z \mathbf{k} \right),
\end{equation}
where $\theta \in [0, \pi]$ represents the angle of inclination between the (laboratory) vertical basis axis $\mathbf{k}_{\text{lab}}$ and the local base-stacking normal vector $\mathbf{e}_3$. The vector $\hat{\mathbf{u}} = (u_x, u_y, u_z)$ is the unit rotation axis directed perpendicular to the plane spanned by $\mathbf{k}_{\text{lab}}$ and $\mathbf{e}_3$, while $\mathbf{i}$, $\mathbf{j}$, and $\mathbf{k}$ represent the imaginary units of the quaternion algebra ($\mathbf{i}^2 = \mathbf{j}^2 = \mathbf{k}^2 = \mathbf{i}\mathbf{j}\mathbf{k} = -1$).

To isolate the local orientation coordinates directly from measured structural data without computing intermediate rotational geometric properties, the components of the orthonormal base triad vectors can also be mapped explicitly onto the coefficients in Eq.~(\ref{quaternioneq}). Let the local normal and in-plane triad axes be given by their Cartesian components in the laboratory frame, where $\mathbf{e}_3 = (e_{3x}, e_{3y}, e_{3z})$, $\mathbf{e}_1 = (e_{1x}, e_{1y}, e_{1z})$, and $\mathbf{e}_2 = (e_{2x}, e_{2y}, e_{2z})$. The unique unit quaternion $\mathbf{q}$ describing this instantaneous local molecular frame is then given by
\begin{equation}
\mathbf{q} = \frac{1}{2\sqrt{1 + e_{3z}}} \left[ (1 + e_{3z}) + (-e_{3y}) \mathbf{i} + (e_{3x}) \mathbf{j} + (e_{1y} - e_{2x}) \mathbf{k} \right]
\end{equation}
where $e_{3x}$, $e_{3y}$, and $e_{3z}$ represent the spatial projections of the base-stacking normal vector, describing the tilt of the nucleotide base plane. The final term along the $\mathbf{k}$-component involves the scalar difference $(e_{1y} - e_{2x})$, which tracks the local internal azimuthal rotation, or twisting angle, of the in-plane vectors $\mathbf{e}_1$ and $\mathbf{e}_2$ around the local stacking axis. This algebraic coupling connects the physical triad positions with the quaternion description.

\subsection*{Discretised computation}
In practice, the continuous parallel transport equation needs to be discretised across sequential discrete nucleotide positions. Let $B_i = [\mathbf{n}_i \mid \mathbf{b}_i \mid \mathbf{t}_i]$ be the $3 \times 3$ orthogonal matrix tracking the Bishop frame at base $i$. To propagate this frame to step $i+1$, the tangent vector field evolves from $\mathbf{t}_i$ to $\mathbf{t}_{i+1}$. The minimal rigid rotation matrix $\Omega_i$ required to align $\mathbf{t}_i$ with $\mathbf{t}_{i+1}$ is evaluated via Rodrigues rotation formula:
\begin{equation}
\Omega_i = \mathbf{I} + \sin(\theta_i) \mathbf{K}_i + (1 - \cos(\theta_i)) \mathbf{K}_i^2
\end{equation}
where the rotation angle is defined by the inner product $\theta_i = \arccos(\mathbf{t}_i \cdot \mathbf{t}_{i+1})$, and $\mathbf{K}_i$ represents the skew-symmetric cross-product matrix generated by the unit rotation axis $\hat{\mathbf{v}}_i = (\mathbf{t}_i \times \mathbf{t}_{i+1}) / \|\mathbf{t}_i \times \mathbf{t}_{i+1}\|$:
\begin{equation}
\mathbf{K}_i = \begin{bmatrix} 0 & -\hat{v}_{z,i} & \hat{v}_{y,i} \\ \hat{v}_{z,i} & 0 & -\hat{v}_{x,i} \\ -\hat{v}_{y,i} & \hat{v}_{x,i} & 0 \end{bmatrix}
\end{equation}
The discretised parallel-transport step update is then resolved by the linear mapping
\begin{equation}
B_{i+1} = \Omega_i B_i
\end{equation}

\subsection*{Triad projection to isolate flip and stacking defects}
Let $R_{i,\text{raw}} = [\mathbf{e}_{1,i} \mid \mathbf{e}_{2,i} \mid \mathbf{e}_{3,i}]$ be the matrix of raw base triad vectors evaluated in the laboratory reference coordinate system. To isolate purely local conformational deformations from global 
structural bending or writhe, the raw base axes are dynamically projected into the moving reference frame generated by the parallel-transported backbone vectors:
\begin{equation}
R_{i,\text{corr}} = B_i^T R_{i,\text{raw}} = \begin{bmatrix} \mathbf{n}_i \cdot \mathbf{e}_{1,i} & \mathbf{n}_i \cdot \mathbf{e}_{2,i} & \mathbf{n}_i \cdot \mathbf{e}_{3,i} \\ \mathbf{b}_i \cdot \mathbf{e}_{1,i} & \mathbf{b}_i \cdot \mathbf{e}_{2,i} & \mathbf{b}_i \cdot \mathbf{e}_{3,i} \\ \mathbf{t}_i \cdot \mathbf{e}_{1,i} & \mathbf{t}_i \cdot \mathbf{e}_{2,i} & \mathbf{t}_i \cdot \mathbf{e}_{3,i} \end{bmatrix}
\end{equation}
Because the transformation matrix $B_i$ satisfies the orthogonality conditions $B_i^T = B_i^{-1}$, this projection represents a rigid spatial coordinate rotation. The column vectors of $R_{i,\text{corr}}$ yield the transformed local base triad directions, denoted as $\mathbf{e}_{k,i}^\text{corr}$. 

The geometric quantities tracking localised flip or stacking defects {\it independent of path curvature} are then calculated as follows,
\begin{equation}
\mathbf{e}_{k,i}^\text{corr} \cdot \mathbf{e}_{k,i+1}^\text{corr} = \left(B_i^T \mathbf{e}_{k,i}\right) \cdot \left(B_{i+1}^T \mathbf{e}_{k,i+1}\right)
\end{equation}
A strong local orientation deformation (such as a base flip or sharp rolling step defect) triggers a localised drop in $\mathbf{e}_{k,i}^\text{corr} \cdot \mathbf{e}_{k,i+1}^\text{corr}$, indicating which molecular axis responsible for the phase singularity.

\section*{GROMACS simulations of DNA with elastic deformations and defects}

Regular $14$-mer B-DNA consisting of $6$ central polyA, polyAT, polyC, and polyCG sequences surrounded by $4$ CG base pair on each side were generated using a biased version of cgDNA+~\cite{shar23}.
List of sequences in table \ref{tab:sequences}. Biasing potentials were applied to the three parameters discussed in the main text: Propeller Twist, Buckle and Opening. Here, we biased cgDNA+ to produces input configurations that are locally deformed and are expected to be close to those later generated during the biased all-atom runs. Structures were later generated using x3DNA software. During this phase, a harmonic bias potential was imposed on the selected parameter $\alpha$ (either Propeller Twist, Opening or Buckle), according to
\begin{equation}
    V_{\alpha}=\frac{K}{2}(\alpha - \bar{\alpha})^2
\end{equation}
where $K=1000$ $\frac{KJ}{mol}$ is the force constant and $\bar{\alpha}$ is the target value of the collective variable.

\begin{table}[ht]
    \centering
    \begin{tabular}{|c|c|}
     \hline
       Name & Sequence\\ \hline
        Seq1 PolyA & {\tt{CGCGAA}}\textbf{AA}{\tt{AACGCG}}\\ \hline
        Seq2 PolyAT & {\tt{CGCGAT}}\textbf{AT}{\tt{ATGCCG}}\\ \hline
        Seq3 PolyCG& {\tt{CGCGCG}}\textbf{CG}{\tt{CGCGCG}} \\ \hline
        Seq4 PolyC & {\tt{CGCGCC}}\textbf{CC}{\tt{CCCGCG}}\\ \hline
        \end{tabular}
    \caption{List of sequences simulated. Bias potential is applied on the two central nucleotides shown in bold.}
    \label{tab:sequences}
\end{table}

\subsection*{All-Atom System Preparation}

Each system was subsequently equilibrated following the standard molecular dynamics (MD) protocol described in~\cite{fass26}, here briefly summarized. The biasing procedure is done using version 2022.5 GROMACS and version 2.8.3 of PLUMED. The force field used is Amber99 parmbsc1~\cite{pere07,ivan16}. The $14$-mer sequences were placed in a dodecahedral box, with 2.0 nm of extra space at the sides of the polymer. Periodic boundary conditions were taken into account and TIP-3P water model is used.~\cite{jorg83}. Particle-Mesh-ewald (PME)~\cite{dard93} model regulates electrostatics, non bonded interactions are screened with a cutoff of $1$ nm. The box is solvated in a $150$ mM NaCl solution, after which the overall charge was neutralized. Leapfrog integrator that includes LINCS constraints is implemented with a time step $\Delta t=2 fs$ to perform an energy minimization that stops at tolerance $1000$ kJ/mol. Later, it is performed a $100$ ps NVT equilibration to fix the temperature at $300$ K, and lastly a $100$ ps NPT simulations, using a Parrinello-Rahman barostat to reach $1.0$ bar. 

\begin{table}[ht]
    \centering
    \begin{tabular}{|c|c|c|}
     \hline
       Parameter & Target Value $\bar{\alpha}$ [rad] & Force constant K [$kJ/mol$]\\ \hline
        Propeller Twist & 0.30 & 300\\ \hline
        Opening & 0.42& 300 \\ \hline
        Buckle & 0.3 & 700\\ \hline
        \end{tabular}
    \caption{List of biased parameters introduced in RBB-NA all-atom MD simulations.}
    \label{tab:biasing_scheme_rbbna}
\end{table}

\subsection*{RBB-NA: biasing all-atom simulations and System relaxation}

Following equilibration, the dodecamers were subjected to enhanced sampling using the RBB-NA algorithm~\cite{voor23}. RBB-NA was initiated from the configurations generated via cgDNA+ procedure. For the selected value of the collective variable, reported in the table \ref{tab:biasing_scheme_rbbna}, MD simulations of $1$ ns were performed using a time step of $2$ fs.
The target value $\bar{\alpha}$, reported in the table, varied for the specific biased parameters but it remains the same across the $4$ different sequences. In all the cases, we chose a weaker $k$ harmonic restraint. This choice was necessary to avoid LINCS constraint instabilities that can arise under excessively strong biasing forces and may lead to an interruption in the simulation. The end point of the biased simulation is then used as a starting point for a $20$ ns unbiased simulation with a time step $\Delta t=2$ fs, whose results are studied in the main text.

\end{document}